# Designing Bugs or Doing Another Project: Effects on Secondary Students' Self-Beliefs in Computer Science

Luis Morales-Navarro, University of Pennsylvania, luismn@upenn.edu
Deborah A. Fields, Utah State University, deborah.fields@usu.edu
Michael Giang, California State Polytechnic University, Pomona, mtgiang@cpp.edu
Yasmin B. Kafai, University of Pennsylvania, kafai@upenn.edu

**Abstract:** Debugging—finding and fixing bugs in code—is a heterogeneous process that shapes novice learners' self-beliefs and motivation in computing. Our Debugging by Design intervention (DbD) provocatively puts students in control over bugs by having them collaborate on designing creative buggy projects during an electronic textiles unit in an introductory computing course. We implemented DbD virtually in eight classrooms with two teachers in public schools with historically marginalized populations, using a quasi-experimental design. Data from this study included post-activity results from a validated survey instrument (N=144). For all students, project completion correlated with increased computer science creative expression and e-textiles coding self-efficacy. In the comparison classes, project completion correlated with reduced programming anxiety, problem-solving competency beliefs, and programming self-concept. In DbD classes, project completion is uniquely correlated with increased fascination with design and programming growth mindset. In the discussion, we consider the relative benefits of DbD versus other open-ended projects.

## Introduction and Background

Debugging—finding and fixing bugs in code—is an essential computational practice, a heterogeneous and open-ended process that shapes how novice learners perceive themselves in relation to computing which impacts their motivation to persist (DeLiema et al., 2022). While debugging is often difficult to learn and to teach (McCauley et al., 2008), encountering bugs can generate fear and anxiety, leading to disengagement and the avoidance of computer science (CS) (Kinnunen & Simon, 2010). Typical debugging teaching practices and curricula emphasize top-down instructionist designs that focus on small, isolated problems and linear strategies for finding well-defined bugs (see McCauley et al., 2008). Contrastingly, in our approach of Debugging by Design (DbD) students create, exchange, and solve buggy open-ended, personally relevant projects. We build on a longstanding tradition of constructionism, emphasizing learner agency by designing applications for others (Harel & Papert, 1990). DbD aims to put students in control over bugs, framing failure as a productive, social experience rather than a negative, discouraging one. Exploratory work on DbD found that students engaged in practices that characterize growth mindsets such as choosing challenges that led to more learning, praising effort, approaching learning as constant improvement, and developing comfort with failure (Morales-Navarro et al., 2021). Weeks after completing DbD, students expressed greater comfort and improved skills in debugging (Fields et al., 2021).

When encountering bugs in computing, students engage with a wide set of values and processes, including self-beliefs that are shaped by experiencing failure and that impact how they react to future failures. Self-beliefs are an array of different self-terms—for instance, self-concept and self-efficacy—that share an emphasis on the beliefs individuals hold about their own abilities and attributes (Valentine et al., 2004), in this case about computing. Measuring student beliefs is important because pernicious belief systems, together with structural inequities, play a role in limiting the participation of historically marginalized groups in computing (Margolis et al., 2017). At the same time, addressing students' beliefs about their own abilities in computing can have an impact in their participation and help shift in their views of CS ability from something that is innate to something that is developed with experience and practice. Since DbD aims to empower students in debugging—by designing creative, multimodal buggy projects for others to solve—we measured student self-beliefs about computing in a quasi-experimental design study within a broader e-textiles unit, asking: Did participation in the DbD/comparison (Music) activity impact students' project completion? Did completing either class project relate to students' CS self-beliefs and was this influenced by whether students were in the DbD or the comparison (Music) activities?

## Debugging By Design

DbD took place within the e-textiles unit of *Exploring Computer Science* (ECS), an inquiry-based CS curriculum committed to broadening participation in computing by addressing the structural inequities and beliefs systems that limit participation from historically marginalized groups (Margolis et al., 2017). During the 10-12 weeks e-textiles unit students create four projects. In these open-ended physical computing activities that integrate coding, circuitry, and crafting, facing failure and unanticipated challenges are expected and inherent in the learning process with problems distributed across modalities (Kafai et al., 2019).

In designing DbD, we extended constructionist approaches by shifting the focus from designing functional artifacts (Harel & Papert, 1990) to designing non-functional, or buggy, projects for others (see Fields et al., 2021). The DbD activity was designed to take place during eight 50-minute-long lessons. Due to the Covid-19 pandemic, together with one experienced e-textiles teacher (not in this study), we created a version of the curriculum that could be taught online with students at home. At the beginning of the activity, students discussed with each other different errors and problems they had encountered when creating e-textile projects. Then student groups decided on the bugs they wanted to include in their designs. They were required to have at least 5 bugs in their code but limited to 1 bug in their circuit diagram. After receiving teacher approval on designs, they wrote a project statement, made a circuit and aesthetic diagram of a buggy project and prepared buggy code for the project (see Figure 1). Teachers then helped students exchange project plans with each other; students built and solved each other's buggy designs. In this study we also designed a *comparison activity* where in place of DbD, students created a new project with programmed music. This provided opportunities to go deeper into programming (by coding tones and rhythms, using arrays, for loops, and conditionals) and the inevitable debugging that happens in creating e-textiles projects without the explicit focus on designing bugs for others, unique to DbD. In this study we were curious to find out the potential benefits of the different activities.

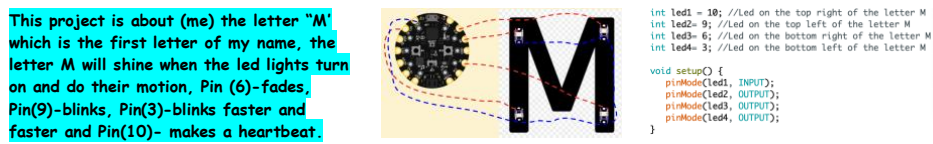

**Figure 1.** A student DbD design: project description, circuitry and aesthetic design, and buggy code excerpt.

## Methods

### Context and Participants

Two experienced teachers at different schools, with high percentages of historically marginalized secondary student populations (58-95% free and reduced lunch; 85-99% non-white), taught eight ECS classes with e-textiles. In Spring 2021 these two schools, in two large West Coast districts, offered online instruction due to the COVID-19 pandemic. Using a quasi-experimental design, five classes participated in the DbD activity (158 students; 90 took the post-survey with 37% identifying as female, 91% speaking a language other than English at home, 58% with no prior CS experience) while three classes (93 students; 54 took the post-survey with 22% identifying as female, 83% speaking a language other than English at home, 46% with no prior CS experience) focused on the comparison (Music) activity. Teachers randomly and pragmatically (e.g., fitting their block schedule) chose classes to implement the activities. Because of the timing of the study at the end of the virtual school year in the second year of the pandemic, there was a high degree of attrition in student participation in the schools. Our IRB allowed for collecting anonymous surveys while identifying the teacher and class period. We did not request information on ethnicity because we believed (and the IRB board agreed) that collecting ethnicity alongside gender by classroom would inevitably make students identifiable. Due to the nature of the implementation of the study, i.e., during a pandemic with virtual schooling, simply comparing DbD against the comparison activity was insufficient as some students started the projects but did not finish them. As such, to investigate the relationship between project completion in both DbD and comparison activities and student CS self-beliefs, we included a survey question about project completion—how far they got in the DbD or music projects— designed to allow us to compare survey responses and how much students participated in the DbD or comparison (Music) activities.

### Data Collection and Analysis

To assess the impact of DbD and comparison activities, 144 students completed a survey at the end of their respective activities. The survey contained a previously validated instrument for nine CS self-beliefs constructs (see Table 1; for details about the instrument see Morales-Navarro, in press) using confirmatory factor analysis (with factor loadings average of .742 and range between .491 to .875; reliabilities between .713 and .889). For the present study, each construct also demonstrated good reliability (all Cronbach's alphas ranged between .729 and .918). For each construct, responses were recorded using a four-point Likert scale to encourage greater reflection and avoid neutral or indifferent responses (1=strongly disagree; 4=strongly agree). To determine project completion in both activities, the survey asked, "Please select the option that best describes how far you got in completing the following project"; responses ranged from "Didn't do it" (1) to "Finished" (5). Teachers created assignments that required students to complete our online survey in their school's learning management systems with time to complete the survey during class and as homework within a certain time. To assess the impact of DbD, two sets of analyses were conducted. First, an analysis of variance (ANOVA) was conducted to examine differences between DbD and comparison activities on project completion. Second, within DbD and comparison activities, correlation analysis examined the potential bidirectional relationship project completion had with the nine CS Beliefs (see Table 1).

## Findings

Notably, ANOVA results found no significant difference between DbD (M = 1.537, SD 2.56) and comparison (M = 2.56, SD = 1.550) activities in terms of project completion, $F(1, 142) = 0.000$, $p = 1$, partial $\eta2 = 0$.

For students doing either DbD or comparison activities, correlation results show that the extent to which they completed the project was related to significant increases in CS creative expression and e-textiles coding self-efficacy (see Table 1). This reinforces findings from previous studies (Kafai et al., 2019) that showed gains in creativity and e-textiles coding self-efficacy during the e-textiles unit without DbD or the extra music project activities. That said, it is worth noting that highly engaged students or those with previous CS backgrounds may have been more likely to complete the projects in both activities. Similarly, access to Wi-Fi and other learning resources during virtual pandemic school very likely influenced project completion.

**Table 1.** Correlation between project completion and constructs across DbD (N = 90) and comparison (N = 54).

| Construct | DbD | Comparison | Construct | DbD | Comparison |
|---|---|---|---|---|---|
| **CS Beliefs** | | | **CS Mindset** | | |
| Problem solving competency beliefs | .191 | .364** | Programming fixed mindset | -.032 | -.178 |
| Fascination in design | .257* | .229 | Programming growth mindset | .313** | .230+ |
| Value of CS | .201 | .266 | **CS Outcomes** | | |
| CS creative expression | .284** | .363** | Programming anxiety | .042 | -.330* |
| E-Textiles coding self-efficacy | .321** | .335* | Programming self-concept | .190 | .297* |

+p < .10, *p < .05, **p < .01, ***p<.001

Among those doing the DbD activity, the extent to which they completed their projects (designing intentionally buggy artifacts and solving their peers' buggy projects) correlated with significant increases in fascination in design and programming growth mindsets. This highlights how designing and solving bugs may influence students' beliefs about design and the potential of creating personally meaningful failure or buggy artifacts in empowering learners to design bugs and engage with their designs in novel ways (Fields et al., 2021). The correlation of project completion with programming growth mindset in the DbD condition supports earlier findings that suggest that agency-driven debugging approaches may be particularly well suited to promote growth mindset (Morales-Navarro et al., 2021). It is also worth considering that while the survey was conducted right after students completed the activity, earlier work found a marked difference in students' perception of DbD immediately after completing it and several weeks after, with most students expressing distress and frustration about bugs right after the activity but comfort and competence with bugs weeks after the activity (Fields et al., 2021). For the comparison activity students, project completion correlated with lower programming anxiety, higher problem-solving competency beliefs, and programming self-concept. This is not surprising as the project pushed students into new programming domains (i.e., music, arrays, etc.), suggesting that this helped students gain confidence in their coding abilities.

## Discussion

Our study takes a holistic approach to studying constructionist, CS education interventions by looking at the relationship between secondary students' computing self-beliefs and completing one of two creative physical computing projects in a quasi-experimental design. One highlight from our results is that students' perspectives on themselves with computing relate to how far they got in making the projects. This finding, across both DbD and comparison groups, foregrounds that making personally relevant e-textiles projects, whether a normal open-ended project or a specifically buggy project, may be beneficial for students' development of self-beliefs in CS. This suggests that progressing toward completion in projects is important for students to fully benefit from constructionist learning activities. As Harel and Papert (1990) argue, creating projects contributes to the affective side of cognition as learners shape their relationships with the concepts they encounter through personal appropriation of knowledge. Even amid a global pandemic at the end of a year of virtual classes, in public schools with high percentages of marginalized students (by class and ethnicity), significant results demonstrate the potential gains from engaging students in making these creative, challenging projects. With Debugging by Design (DbD), project completion was uniquely correlated with programming growth mindset, demonstrating a potentially distinct benefit of DbD that complements open-ended, constructionist-driven projects more generally. These findings are important because programming growth mindset is particularly important for retention, perseverance, and endurance amongst marginalized populations in CS (Kinnunen & Simon, 2010; Margolis et al., 2017). Therefore, educators and researchers should further consider the role that integrating DbD projects may have in supporting students when making projects.

Our research provides many potential directions for future study. Research should explore the specific roles of different aspects of DbD—both designing bugs for peers and solving bugs by peers—in fostering growth mindsets. Our earlier case study research in one classroom documented that students exhibited growth mindset practices largely while designing bugs (Morales-Navarro et al., 2021), yet this current study demonstrates the importance of doing the entire scope of the DbD project, from designing to solving peers' buggy projects. At the same time, replicating this study in in-person classroom settings and conducting case studies of students' design processes could be beneficial to better understand the differences between DbD and comparison activities in relation to self-beliefs without so many other factors in play (e.g., a pandemic, virtual education). Further, while we attend to self-beliefs in this study, future research should also investigate what contributions DbD or comparison activities make to learners' conceptual understanding, for instance developing breadth of knowledge about types of bugs or strategies for identifying and solving bugs. Finally, we have explored DbD in one specific context—designing e-textiles in an introductory computing course; many other applications are possible.

## Acknowledgments

This work was supported by a grant from the NSF to Yasmin Kafai (#1742140). Any opinions, findings, and conclusions or recommendations expressed in this paper are those of the authors and do not necessarily reflect the views of NSF, the University of Pennsylvania, Utah State University, or California State Polytechnic University.